\documentclass{sig-alternate}
\begin{document}
\title{Multi-mode Sampling Period Selection\\
for Embedded Real Time Control}
\author{
Rajorshee Raha, Soumyajit Dey, Partha Pratim Chakrabarti, Pallab Dasgupta\\
Department of Computer Science \& Engineering\\ Indian Institute of Technology
Kharagpur, INDIA\\
\{rajorshee.raha, soumya, ppchak, pallab\}@cse.iitkgp.ernet.in}
\maketitle
\begin{abstract}
Recent studies have shown that adaptively regulating the sampling 
rate results in significant reduction in computational resources in
embedded software based control. Selecting a uniform sampling rate
for a control loop is robust, but overtly pessimistic for sharing processors
among multiple control loops. Fine grained regulation of periodicity achieves
better resource utilization, but is hard to implement online in a robust way.
In this paper we propose multi-mode sampling period selection, derived from an offline control
theoretic analysis of the system. We report significant gains in
computational efficiency without trading off control performance.
\end{abstract}
%
%
\terms{Real Time Control, Automotive Software}
\section{Introduction}
Embedded software-based control systems have traditionally been implemented by assuming fixed 
sampling rates and fixed task periods~\cite{Buttazzo}. The sampling rate is derived from a control theoretic
analysis~\cite{seto} of the system in a manner that guarantees desired level of control performance at all
reachable states of the system. A uniform period can be implemented robustly, since we can 
analyze the sampling rate of all the control tasks and then choose an appropriate computational
infrastructure that can statically schedule a periodic execution of the software components
of these control tasks. The schedule does not change during execution and hence the control 
performance is deterministic.
\par Recent studies confirm a widely accepted belief, namely that a uniform sampling rate is not
a good choice when multiple control loops share a common computing resource such as an 
Electronic control unit (ECU). These studies establish that the sampling period can be 
regulated to achieve significant benefits in computational performance without any trade off
in control performance. In fact, it has also been shown that a non-uniform scheduling strategy
can balance the sampling rates among the control loops sharing a ECU in such a way that the 
overall control performance improves~\cite{cervin2011paper,henrikson}.
\par Uniform sampling rate is typically a pessimistic choice, since we need to choose a sampling
rate that guarantees control performance at all reachable states of the system. It is often
the case that the selected rate is necessary at only specific control states of the system,
whereas at all other states a much lesser sampling rate suffices. Adaptive sampling derives
its benefit from this fact by intelligently regulating the sampling period as needed to
maintain the desired level of control performance~\cite{cervin2011paper}.
\par Fine grained regulation of the sampling rate may theoretically determine the optimal balance
between computational efficiency and control performance but such schemes are difficult to
implement in practice due to non-determinism in timing introduced by the computational 
infrastructure (including message delays, execution time variations in different paths
of the control software, etc). If the sampling rate is known a priori then it becomes
possible to develop an appropriate schedule for all control tasks sharing a ECU with
adequate consideration for these types of non-determinism.
\par In this paper we profess the use of coarse grained regulation of the sampling rate. 
Specifically we propose an approach where for each control loop, a limited number of 
sampling rates are chosen and a control theoretic analysis is used to determine the 
switching criteria between these modes. Therefore, for each control loop we have a
set of {\em sampling states}, and an automaton that captures the switching between
these sampling states. The {\em global sampling state} of the system is a concatenation
of the sampling states of the control loops in the system. For each global sampling
state, the exact schedule for the control tasks is precomputed considering all
types of non-determinism arising out of the execution of software tasks. The reachable
global sampling states are chosen based on available computational bandwidth and
the relative priorities of the control loops.
\par The primary objective of this paper is to establish the benefit of using multiple discretely chosen sampling rates. In
the process, we also present the following enabling contributions.
\begin{enumerate}
\item We outline the basis for choosing the various sampling rates based on use case analysis.
\item We present an analytical approach which determines the criteria for switching between sampling rates so that control performance is not hampered.
\item We present the construction of an automaton based scheduler, which implements the switching between sampling rates of the controller.
\end{enumerate}
We present the necessary background for the work in Section \ref{priliminary}. We outline the methodology of multi-mode
sampling period selection in Section \ref{multimode}. Subsequently, we use the control theoretic model of an Anti-lock
Braking System (ABS) as a running example to validate our approach. 
\section{Background Study}\label{priliminary}
\begin{figure}[h]
 \centering \vspace{-.2cm}
 \epsfig{file=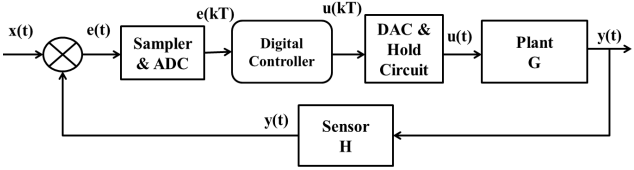, height=.9in, width=3.3in}\vspace{-.4cm}
 \caption{Discrete Time Control System}\vspace{-.3cm}
    \label{fig:Digital Controller}
\end{figure}

In this section, we outline the mathematical relation between the sampling period of a discrete time controller
and its control stability. Any discrete time feedback control system can be
represented as shown in Fig. \ref{fig:Digital Controller}~\cite{discretecontrolbook}, here $x(t)$ is the input
to the system, $e(t)$ is the error signal, $u(t)$ is the controller output and $y(t)$ is the plant output fed back to
the controller using a sensor.
\par Generally, the sampling of the continuous signal is done at a constant rate $T$, which is known as the sampling period
or interval. The sampled signal $e_k=e(kT)$, is the discretized signal with $k\in\mathbb{N}$. In general, for any transfer function,
the control signal output depends on $n$ previous control signal output instances and $m$ previous error signal
instances~\cite{discretecontrolbook}. The situation may be represented as,\vspace{-0.1cm}
\begin{equation}
 \label{eq:digital control signal generalized 1}
  \begin{split}
    u_k &= -a_1u_{k-1} -a_2u_{k-2} - \dots - a_nu_{k-n} + b_0e_{k} +\\
	& b_1e_{k-1} + b_2e_{k-2} + \dots + b_me_{k-m}\\
  \end{split}
\end{equation}
\par The discrete signals $u_{k-1}$,$u_{k-2},..$ are the delayed versions 
of $u_k$ by sampling period $T, 2T,..$ respectively. In Laplace domain, a time delay is introduced into a signal by
multiplying its Laplace transform by the operator $e^{-T s}$. Let the Laplace domain representation of $u_k$ and $e_k$ 
be $U(s)$ and $E(s)$ respectively~\cite{kuo_book}. Hence, $u_{k-1},u_{k-2},..$ can be represented in Laplace 
frequency domain as $e^{-Ts}U(s),$ $e^{-2Ts}U(s)$ $,..$ respectively and similarly for $e_{k-1},..$ . Thus, the
corresponding Laplace domain representation of Eq. \ref{eq:digital control signal generalized 1} shall be,
\begin{equation}
 \label{eq:laplace}
\begin{split}
U(s) &= - a_1e^{-Ts}U(s) - a_2e^{-2Ts}U(s) - \dots + b_0E(s) +\\
& b_1e^{-Ts}E(s) + b_2e^{-2Ts}E(s) + \dots \\
\end{split}
\end{equation}
\par Substituting $e^{Ts}$ with the discrete frequency domain operator $z$~\cite{discretecontrolbook} and simplifying this further 
we get the discrete time transfer function of the controller $C(z)$ as,
\begin{equation}
 \label{eq:TF discrete 2}
\frac{b_0z^n+b_1z^{n-1}+..+b_mz^{n-m}}{z^n+a_1z^{n-1}+..+a_n}=b_0\frac{\Pi_{j=1}^m(z-z_j)}{\Pi_{i=1}^n(z-p_i)}z^{n-m}
\end{equation}
where $z_j$ are the zeros and $p_i$ are the poles of the transfer function. Behavior of any discrete time controller
can be observed by analyzing the poles and zeros of the corresponding transfer function~\cite{discretecontrolbook}. The 
positions of the poles and  zeros differ for different sampling intervals $(T)$. Correspondingly, the control stability
of the overall system gets effected. More related background is provided in Appendix \ref{A:Control}.
\section{Methodology Outline}\label{multimode}
In this section, we outline our proposed methodology of multi-mode sampling period selection for embedded real time control.
The main steps of this proposed multi-mode methodology are as follows,
\begin{itemize}
 \item Step I : Developing a control theoretic model of the corresponding system.
 \item Step II : Classification of different modes based on different control parameters and selection 
of best possible sampling rate for the corresponding operating modes. 
 \item Step III : Construction of a supervisory automaton for controlling the mode switching.
\end{itemize}
\begin{figure}[h]
 \centering \vspace{-.3cm}
 \epsfig{file=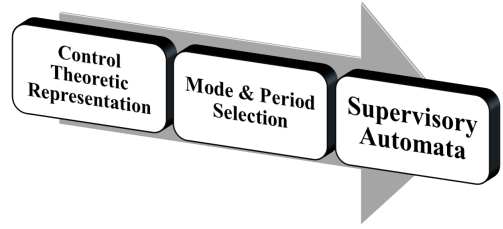, scale=0.2} \vspace{-.4cm}
 \caption{Methodology Outline} \vspace{-.25cm}
    \label{fig:outline}
\end{figure}
\par In the following sections we demonstrate the proposed methodology with an extensive analysis of ABS as 
a running example. In Section \ref{ABS_model}, we present the control model of the ABS.
In Section \ref{multi-mode}, we divide the driving
pattern of a vehicle into multiple modes parameterized by vehicular characteristics like velocity, brake pedal pressure and slip. For each 
such mode, we choose a sampling frequency which ensures stability guarantee of the ABS. In Section \ref{automata}, we
outline our approach for guard condition selection for switching between different modes, and synthesize a scheduler automaton which
may supervise the mode selection depending on vehicular dynamics.
In Section \ref{final-results}, we provide experimental results supporting the proposed 
approach.
\section{Control Model}\label{ABS_model}
\begin{figure}
 \centering
 \epsfig{file=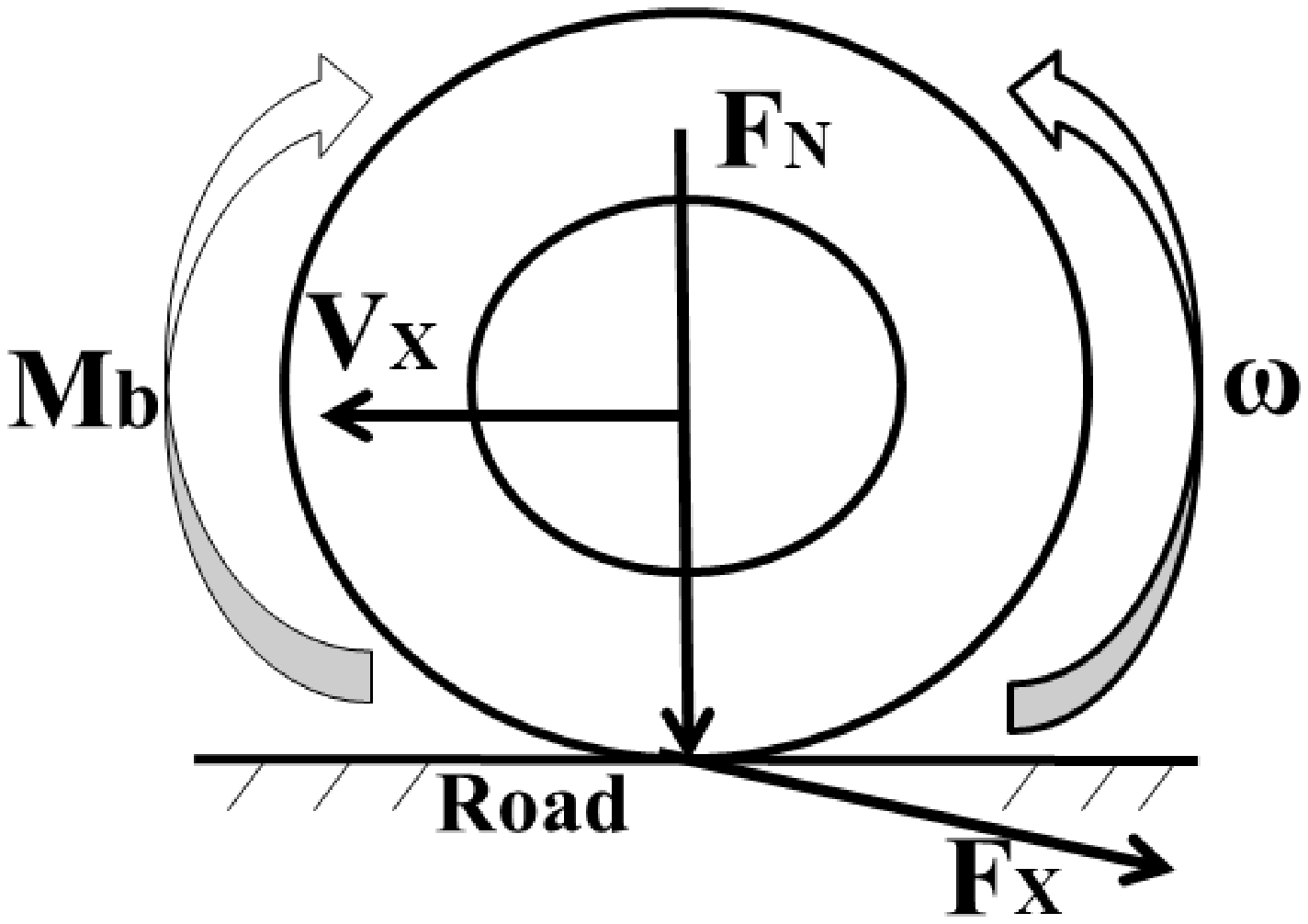, scale=0.25} \vspace{-.4cm}
 \caption{1/4 Car Forces and Torques~\cite{bosch_safety}} \vspace{-.5cm}
\label{fig:car}
\end{figure}

\begin{figure}
 \centering \vspace{-.4cm}
  \epsfig{file=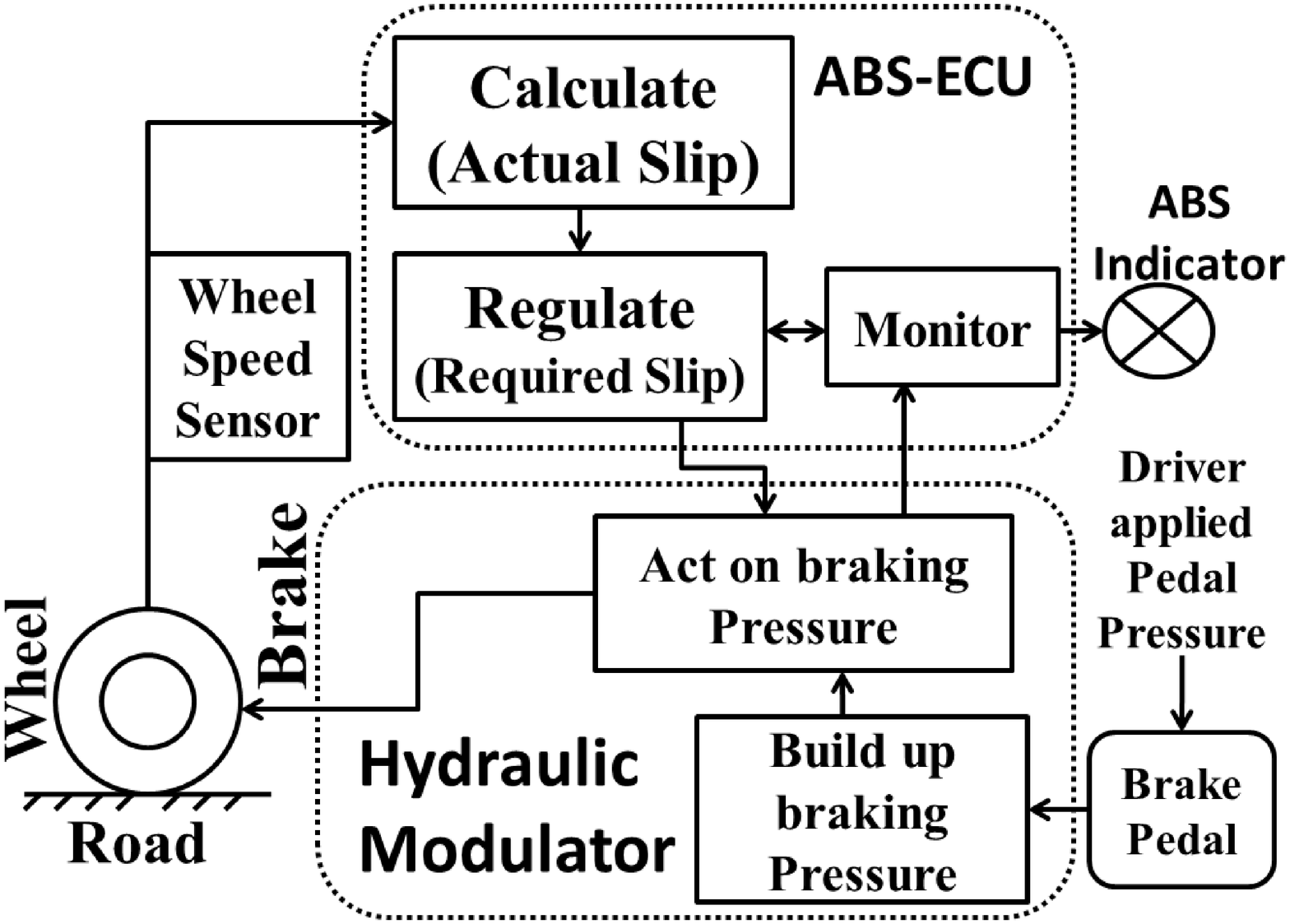, height=1.2in, width=2.5in} \vspace{-.4cm}
  \caption{ABS Overview~\cite{bosch_safety}}
    \label{fig:ABS_Controller}
\end{figure}

ABS is an automobile safety critical driver assistance system which prevents the wheels
from locking and avoids uncontrolled skidding.
An abstract block diagram of a vehicle with ABS is shown in Fig. \ref{fig:ABS_Controller}.
For designing the vehicle we used a simplified quarter car model as shown in Fig. \ref{fig:car}.
\par Here, m is the mass of the quarter vehicle, $V_x$ is lateral speed of the vehicle,
$\omega$ is the angular speed of the wheel, $F_N$ is the vehicle vertical force, $F_x$ is the frictional force transmitted to the road,
$M_b$ is the braking torque, R is the wheel radius and $J_\omega$ is the wheel inertia.
Wheel slip $\lambda$ is given as, $\lambda = 1 - \omega R/V_x$.
\par
\begin{figure}[h]
 \centering \vspace{-.25cm}
 \epsfig{file=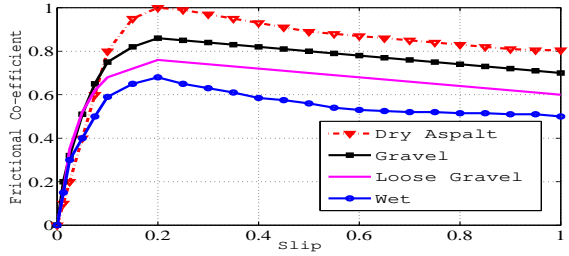, height=1.4in, width=3.4in} \vspace{-.65cm}
 \caption{$\mu-\lambda$ Curve~\cite{wong2001theory}} \vspace{-.2cm}
\label{fig:mu-slip}
\end{figure}

 The effective braking force is dependent on the frictional force~\cite{bosch_safety} transmitted to the road which is related to $F_N$ as,
$F_x$ = -$\mu F_N$, where $\mu$ is the frictional coefficient of the road surface. Thus, as evident
from Fig. \ref{fig:mu-slip}, the amount of slip will vary depending on road conditions.
\par Relationship between wheel slip and frictional coefficient can be approximated, $\mu=f(\lambda)$, using a piecewise linear
function~\cite{khalil1992nonlinear} as,
\begin{equation}
 \label{eq:mu linear}
  \mu =
  \begin{cases}
     \alpha \lambda ,& \lambda \leq 0.2 \\
     - \frac{1}{2} \lambda +\frac{3}{4} + \beta, & \lambda > 0.2
  \end{cases}
\end{equation}
where, $\alpha$ $\in$ $[0,8]$ and $\beta\in [-0.1,0.1]$. The non-linear equations for designing a
quarter car model can be given as,
\vspace{-.15cm}\begin{equation} 
 \label{eq:v-dot}
\begin{array}{c}
 \dot{V_x} = -\frac{1}{m}F_N\mu \\
  \dot{\omega} = \frac{R}{J_\omega}F_N\mu - \frac{M_b}{J_\omega}\\
  \dot{\lambda} = -\frac{1}{V_x}[\frac{1}{m}(1-\lambda)+\frac{R^2}{J_\omega}]F_N\mu + \frac{1}{V_x}\frac{R}{J_\omega}M_b
\end{array}
\end{equation}
\par Using Taylor series expansion method~\cite{khalil1992nonlinear,ABS_paper} for linearizing a nonlinear system we obtain a
linear (affine) system description from Eq. \ref{eq:v-dot} as,
\begin{equation}
 \label{eq:ABS linear}
 \begin{split}
    \dot{x} &= A_l x + E_l + B_l u^* \\
     y &= C_l x + D_l u^*
 \end{split} 
\end{equation}
where, $x^T = [V_x,\lambda]$, $u^* = M_b \cdot V_x$ , $y = [\lambda]$,
$A_l$, $B_l$, $C_l$, $D_l$ are system input and output matrices respectively.
$E_l$, $l=f(x)$ are the affine term and function telling the validation of linearizion.
The formation of this state space equation from the nonlinear equations are described in
details in Appendix \ref{A:0}.
The objective of ABS controller is to decelerate the vehicle as fast as possible, while maintaining its steer ability 
by minimizing wheel slip. The main components of ABS are the ABS-ECU, hydraulic
modulator, and wheel speed sensor. The ECU constantly monitors the wheel rotational speed through the wheel speed
sensors, and also measures the actual slip. The controller's task is to maintain the braking torque within a certain range.
The braking force is applied to the wheels by the hydraulic modulator. It rapidly pulses the brakes to prevent
wheel lock up, even during panic braking in extreme conditions and promises shortest possible distance under most conditions.
We can design a discrete PID controller for this purpose as,
\begin{equation}
 \label{eq:ABS_PID}
 \begin{split}
    M_b &= K_p e + K_i \int e dt + K_d \frac{de}{dt} \\
    M_b(z)  &=  [K_p + \frac{K_iT}{z-1} + \frac{K_d(z-1)}{T}]E(z)
 \end{split}
\end{equation}
\par Here $K_p$, $K_i$, $K_d$ are the proportional, integral, and derivative gain respectively of the 
PID controller. $T$ is the sampling interval.
$M_b(z),E(z)$ are the discrete domain representation of the the braking torque, $M_b$ and the error signal, $e=\lambda_d-\lambda$,
which is the difference between the desired slip $(\lambda_d)$ and actual slip $(\lambda$).
Observe that when $\lambda = \lambda_d$ then $e=0$, i.e.
$\dot{\lambda}=0$. Substituting this in Eq. \ref{eq:v-dot}, $M_b$ can be represented as,
\vspace{-.15cm}
\begin{equation}\label{eq:M_b}
  M_b = [(\lambda-1)\frac{J_\omega}{R} -mR]\dot{V_x}
\end{equation}
Hence, it is obvious from Eq. \ref{eq:ABS_PID} $\&$ \ref{eq:M_b} that the control performance, i.e. stability of the
ABS controller will vary for different values of $T$, $V_x$ and $\lambda$.
\section{Mode and Period Selection}\label{multi-mode}
The candidate sampling modes and periods for a controller are determined by partitioning its input
space based on the use-case scenarios and stability of the control law in those scenarios. For 
example, in ABS, the adequacy of a sampling rate in a given scenario depends on the
urgency of braking (which is a function of vehicle speed and pedal pressure) and the slip ratio
(which is a function of the vehicle speed and friction on the road). In general, we select the 
different possible sampling frequencies of the controller using the following approach. 
\begin{enumerate} 
\item Identify vehicular parameters which impact the controller output (i.e. braking torque in this case).
\item Identify multiple possible driving scenarios and corresponding ranges of vehicular parameters and also the probable driver response. 
\item Perform stability analysis followed by identification of maximal acceptable sampling period in each driving scenario. 
\end{enumerate}
\par
\begin{figure}[h]
  \centering \vspace{-.32cm}
\resizebox{\linewidth}{!}{
    {\epsfig{file=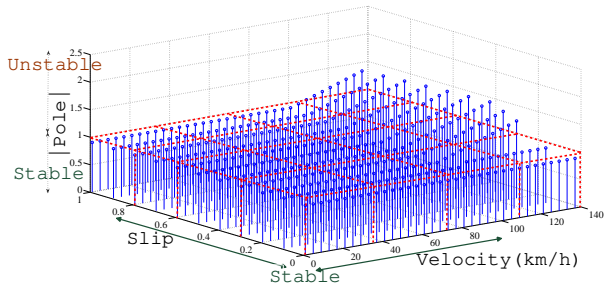}}}
 \vspace{-.8cm} \caption{N0 Mode Stability Guarantee} \vspace{-.23cm}
    \label{fig:ABS Mode-N0}
\end{figure}

 Steps 2 and 3 may have to be iterated to arrive at a gainful combination of sampling modes.
A sampling mode is effective towards gaining computational efficiency only if the controller stays
in that mode for a non-trivial period of time. Therefore it is necessary to relate the sampling
modes with different use-case scenarios. In ABS, we estimate different possible traffic 
scenarios (city traffic, suburban or medium traffic \& highway traffic) and the variation in traffic
density, traffic regulations and corresponding average cruising speed and driver reaction to arrive
at the sampling modes. We categorize the brake pedal pressure range as low, mild, medium and high,
and also consider various speed ranges and slip ratios. For each of the scenarios, we determine the
sampling rate at which the controller is stable. Through this study we selected three sampling modes,
as outlined below:

\noindent $\bullet$ \textbf{N0 Mode:} This mode requires the least sampling rate among the three
chosen mode, and targets scenarios where the vehicle is cruising at low to medium speeds (such as 
in city traffic). Considering average cruising speed and the driver reaction, we arrive at the
operating sampling rate $T_s = 0.2ms$ in which the ABS achieves satisfactory control performance as given by Fig.
\ref{fig:ABS Mode-N0}. For a given velocity (X axis) and slip (Y axis), we carry out the standard unit circle analysis
and plot the maximum magnitude among the different pole positions (Z axis) of the transfer function corresponding to
$T_s = 0.2ms$. The stability variation for different slip values is due to the piecewise linear function given in Eq.
\ref{eq:mu linear}. It may be observed from Fig. \ref{fig:ABS Mode-N0}, all the poles are of magnitude $<=1$ for
velocity range $[0\dots85]km/h$ and slip range $[0\dots0.65]$ thus ensuring stable vehicular dynamics. Thus
correspondingly brake pedal pressure variation range is selected as $[low,mild,medium]$ estimating the probability of
the above mentioned slip range. We carry out similar analysis for the other cruising modes and derive satisfactory
sampling intervals.

\begin{figure}[h]
  \centering \vspace{-.32cm}
\resizebox{\linewidth}{!}{
    {\epsfig{file=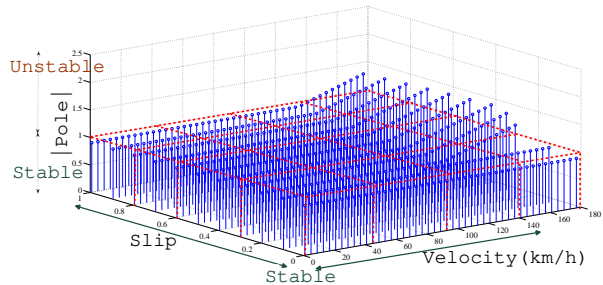}}}
   \vspace{-.8cm} \caption{N1 Mode Stability Guarantee} \vspace{-.23cm}
    \label{fig:ABS Mode-N1} 
\end{figure}

\noindent $\bullet$ \textbf{N1 Mode:} This mode uses higher sampling rate than N0, and targets suburban traffic
scenarios. The maximum sampling interval with stability guarantee is found to be $T_s=0.15ms$ as shown in 
Fig. \ref{fig:ABS Mode-N1}.

\begin{figure}[h]
  \centering \vspace{-.25cm}
\resizebox{\linewidth}{!}{
    {\epsfig{file=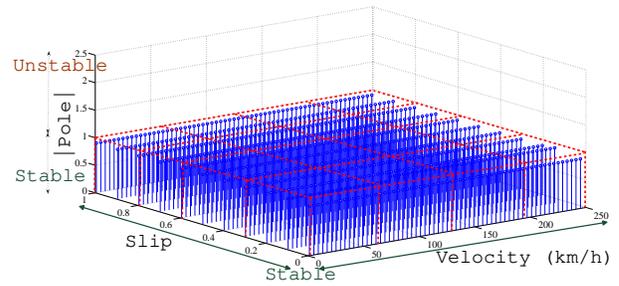}}}
   \vspace{-.8cm} \caption{E Mode Stability Guarantee}\vspace{-.45cm}
    \label{fig:ABS Mode-E} 
\end{figure}

\noindent $\bullet$ \textbf{E Mode}: This mode uses a sampling rate that is adequate in all scenarios. Existing
approaches for choosing a uniform sampling mode will choose this sampling rate. We choose the sampling frequency
for which the vehicle remains stable considering all velocity and brake pedal pressure variations as shown in 
Fig. \ref{fig:ABS Mode-E}. The corresponding sampling period for our model is found to be $T_s=0.1ms$. We 
designate this as \textit{emergency} sampling mode, in case of any driving irregularity the controller switches
to this mode, thus ensuring vehicular stability.
\par Z-domain unit circle stability analysis is used to show the vehicular parameters and stability relationship
graphically for extensive parameter ranges. Similar observations can be obtained mathematically using other nonlinear
system stability~\cite{discretecontrolbook} criteria like, Lyapunov, Nyquist, Routh-Hurwitz or Bode plot analysis.
The effective braking pressure range is varied in each of these modes to achieve satisfactory performance, depending upon
driving scenarios. The mode switching and detailed switching criterion are discussed in the next section.
\section{Supervisory Automata}\label{automata}
\begin{figure}[h]
  \centering \vspace{-.25cm}
    {\epsfig{file=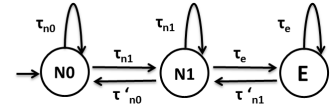, scale=0.25}} \vspace{-.4cm}
    \caption{Scheduler Automata} \vspace{-.15cm}
    \label{fig:automata}
\end{figure}

Our analysis of the different possible driving scenarios and the choice of sampling periods entails the creation of a scheduler 
which may dynamically switch the controller among different sampling modes. Such a supervisory automaton is shown in 
Fig. \ref{fig:automata}. 
The criteria for switching between the three modes of the  automaton is chosen based on our observations about vehicular parameters 
vis-a-vis stability. Let ``$v$'' and ``$bpp$''  be the velocity and brake pedal pressure  respectively. The brake pedal pressure is divided in
to the ranges low, mild, medium and high. High brake pedal pressure  represents the probability of larger value of slip. Medium, mild and low 
brake pedal pressure signifies lesser value of slip. The guard conditions for switching between states are given as,
{\small
\begin{eqnarray*}
\tau_{n0} & = & (v\in[0\dots85]\ \&\ bpp\in[low,mild,medium]) \\
& & \quad \mid (v\in[85\dots140]\ \&\ bpp\in[low,mild]) \\
\tau'_{n0} & = & (v\in[0\dots80]\ \&\ bpp\in[low,mild,medium]) \\
& & \quad \mid (v\in80\dots135]\ \&\ bpp\in[low, mild]) \\
\tau_{n1} & = & (v\in[0\dots85]\ \&\ bpp\in[high])\\
& & \quad \mid (v\in[85\dots140]\ \&\ bpp\in[medium,high]) \\
& & \quad \mid (v\in[>140]\ \&\ bpp\in[low,mild]) \\
\tau'_{n1} & = & (v\in[0\dots80]\ \&\ bpp\in[high]) \\
& & \quad \mid (v\in[80\dots135]\ \&\ bpp\in[medium,high]) \\
& & \quad \mid (v\in[>135]\ \&\ bpp\in[low,mild]) \\
\tau_e & = & (v \in [>140]\ \&\ bpp\in[medium,high])
\end{eqnarray*}
}
The guard conditions for switching between `$N0$' to `$N1$' and vice versa are selected to be `$\tau_{n1}$' and `$\tau'_{n0}$' respectively. 
Similarly, for `$N1$' to `$E$' and vice versa, switching conditions are `$\tau_{e}$' and  `$\tau'_{n1}$' respectively. The guard conditions are 
selected ensuring some amount of hysteresis while mode switching.  For example, the automaton switches from mode `$N0$' to mode `$N1$' in case 
$bpp=high$ and $v\in[0\dots85]$. However, the automaton switches from mode `$N1$' to mode `$N0$' when $v\in[0\dots80]$ and $bpp$ is not $high$.
Similarly, the automaton switches from mode ``$N1$'' to $E$ when $bpp$ is $medium$ or $high$ and $v\in[>140]$, however, the automaton
switches from mode `$E$' to mode `$N1$' when $v\in[>135]$ and $bpp$ is $low$ or $mild$.
\par Whenever the automaton makes a transition from a mode with lower sampling period to a mode with higher sampling period (e.g. $E$ to $N1$), 
it ensures that the guard conditions are valid for a  certain prefixed number of clock cycles.  In that way, unwanted glitches due to faulty sensor readings 
are expected to be filtered out. 
\par The main objective of synthesizing the scheduler automaton was to reduce ECU bandwidth requirement. Scheduling in
$E$ mode signifies a sampling periodicity of $0.1ms$, while the sampling periodicity of $N0$ and $N1$ modes are $0.2ms$ 
and $0.15ms$ respectively. If we notice the mode switching scenarios, we observe that when the car is cruising at a 
certain speed and no brake pedal pressure is applied, the controller is scheduled using infrequent sampling periods
($N0$ or $N1$). Further, in scenarios when the car is cruising at a certain speed and brake pedal pressure is applied,
the mode  switching will be supervised by the respective scheduler automaton ensuring that it will switch to a more
frequent sampling mode. Thereby, in a general cruising scenario, a multi-mode controller ensures nearly $30\%-50\%$ ECU
bandwidth saving as shown in our simulation results.
\section{Results}\label{final-results}
The initial part of this section is devoted towards establishing the motivation for multi-mode sampling through
experimental results. The latter part of the section demonstrates the gain in computational bandwidth and
the benefit of effective sharing of computational resources between multiple controllers.

\par We analyze the performance of the ABS for different sampling intervals. Observing the $\mu$ vs
$\lambda$ curve in Fig. \ref{fig:mu-slip} carefully, we notice that the peak point on most of the road scenarios belong
to the range [0, 0.2]. Hence, for effective braking with maximum possible road friction, we set $\lambda_d = 0.2$ as
the desired slip. In our experimental setup, when brake force is applied with current velocity $V=100km/h$, it is
expected to gradually decrease until $V = 0km/h$ and throughout this deceleration phase the slip value should be as
close as possible to the desired slip ($\lambda_d$ =0.2) thus ensuring smooth braking. Finally the slip value should be
1 (normalized slip) when the vehicle comes to rest. 
\begin{figure}[h]\vspace{-.35cm}
    \hspace{-.4cm}{\epsfig{file=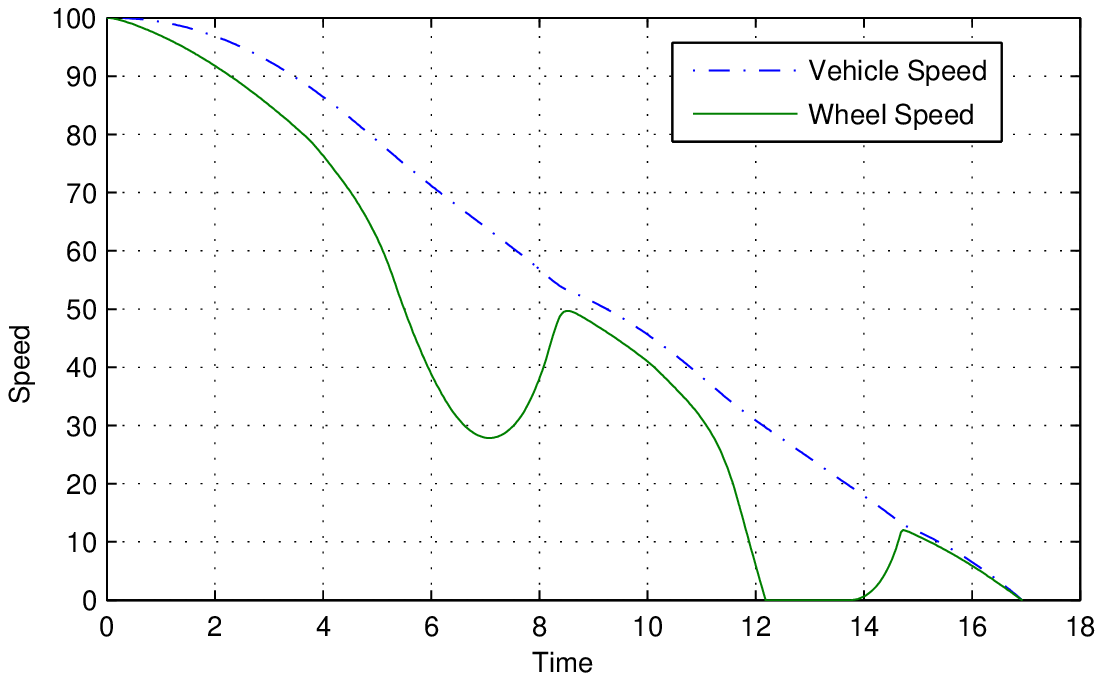, scale=0.36}}
    \hspace{-.4cm}{\epsfig{file=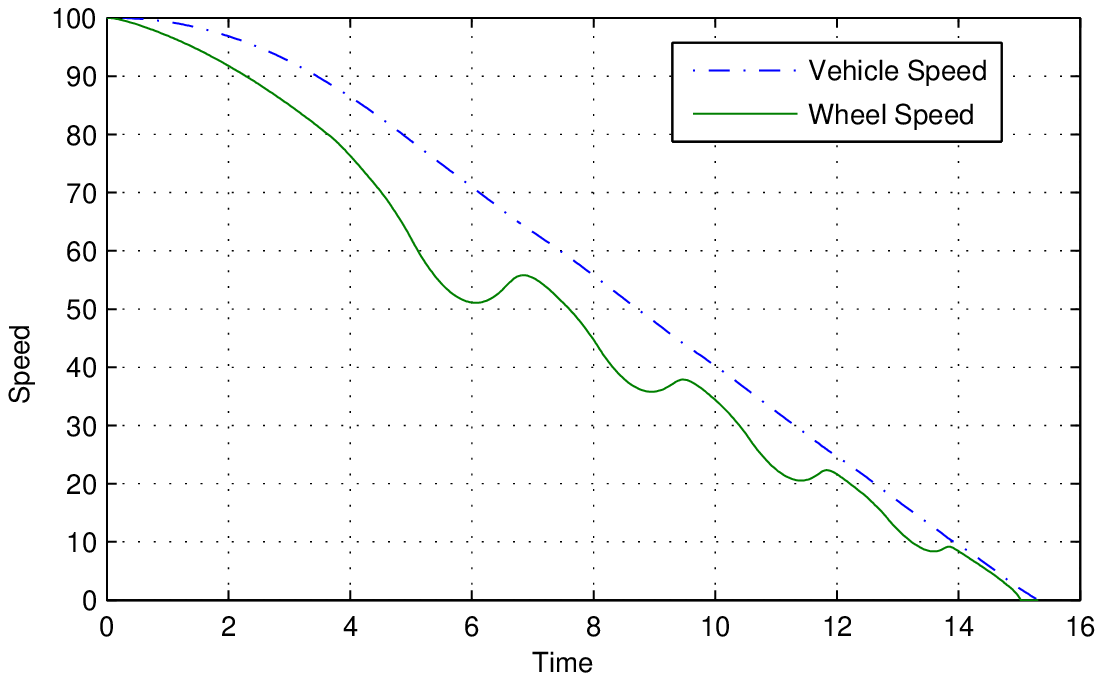, scale=0.36}}
    {\epsfig{file=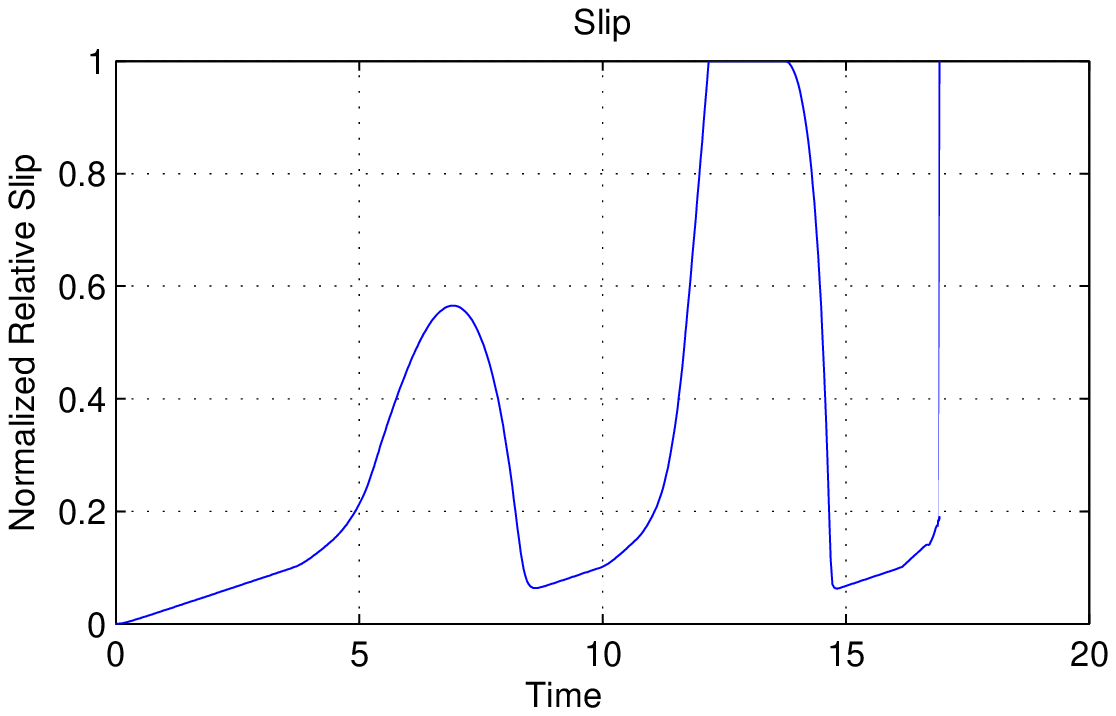, scale=0.32}}\vspace{-.2cm}
    {\epsfig{file=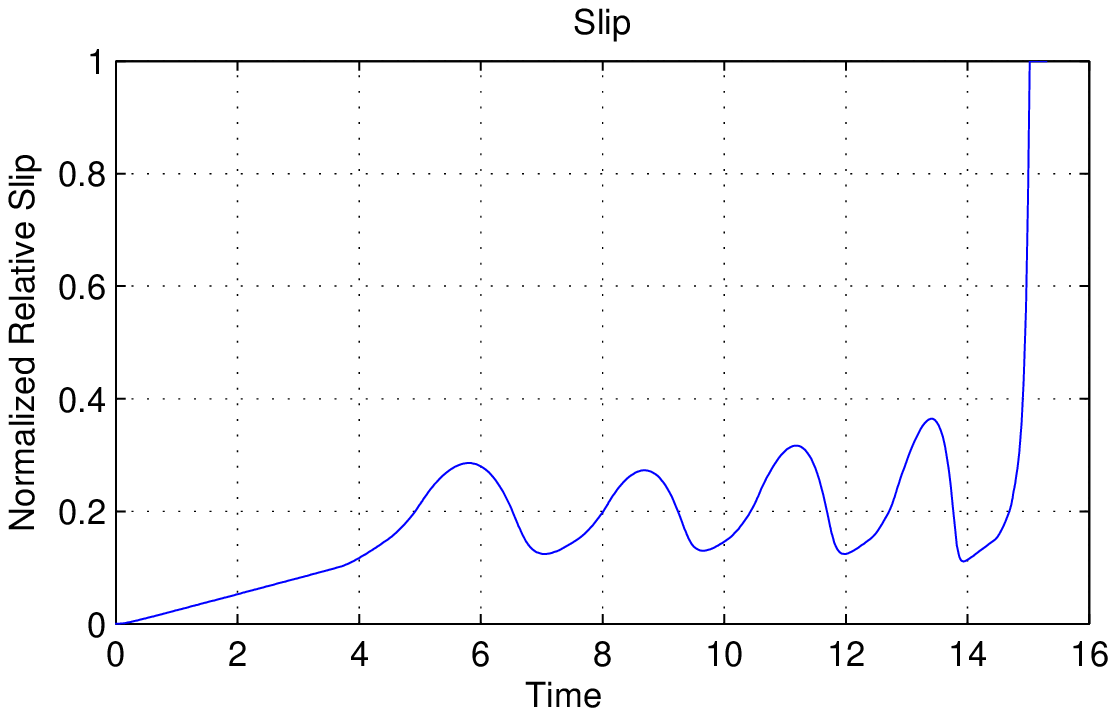, scale=0.32}}\vspace{-.2cm}
    \caption{Slip Variation: the Left \& Right column figures correspond to sampling time $T_{s1}=1s$ \& sampling time
$T_{s1}=0.01s$ respectively.}
    \label{fig:ABS_slip}
\end{figure}
\par We observe that the slip varies significantly for two different sampling rates as given by Fig. \ref{fig:ABS_slip}.
Observe from the left column in Fig. \ref{fig:ABS_slip} that with a choice of moderate sampling rate, the slip varies
drastically thus leading to undesired perturbations in vehicular speed while braking. However, as shown in the right
column of the same figure, with the higher sampling rate, the slip exhibits a well damped trajectory around the desired
value ($\lambda_d$ =0.2) leading to smoother vehicular deceleration. 

\par
\begin{figure}[h]
\vspace{-.35cm}
  \epsfig{file=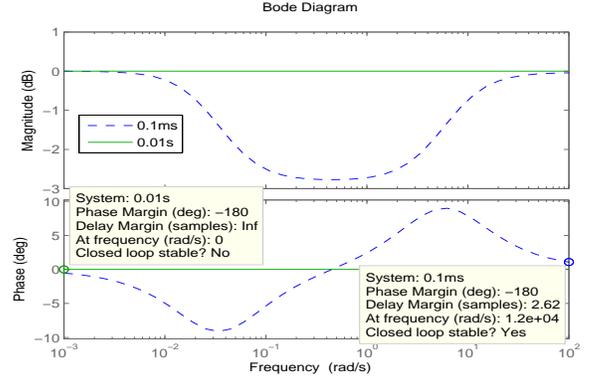, height=2.1in, width=3.5in}\ \vspace{-.6cm}
  \caption{Stability Variation:$T_s=0.01s$, Unstable $\&$ $T_s=0.1ms$, Stable}
    \label{fig:ABS_bode} \vspace{-.2cm}
\end{figure}

 Our notion of control performance is based on ensuring the stability of the system. We empirically
observe the variation of stability with sampling rate. For stability analysis, we used unit circle as well as bode 
plot analysis~\cite{discretecontrolbook}. We take a moderate velocity, say $V = 60km/h$ and calculate the stability of 
the system for different sampling intervals. We show one such example Bode plot for stability analysis in Fig.
\ref{fig:ABS_bode}. It may be observed that with the sampling interval $T_s = 0.01s$, the system is unstable. However,
if we drastically reduce the sampling interval to $0.1ms$, the system becomes stable. In Appendix \ref{A:1} and
\ref{A:2}
more such results are given.

\par
\begin{figure}
  \centering
   \epsfig{file=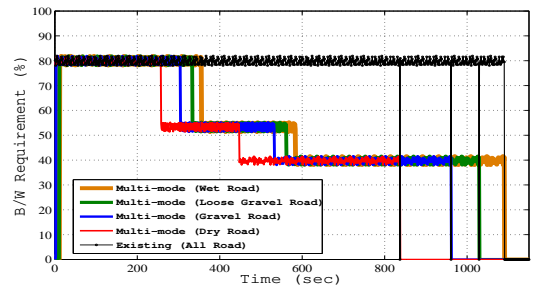, height=1.6in, width=3.2in} \vspace{-.4cm}
    \caption{ECU, Panic Braking Scenario} \vspace{-.5cm}
    \label{fig:ECU_comparision}
\end{figure}

 We have employed our methodology of multi-mode sampling period selection for the ABS running example. We consider
a braking scenario with the initial and final speed being $200km/h$ and $0km/h$ respectively and compare the minimum
stopping distance achieved by our `Multi-mode' ABS Controller with the existing Matlab model of ABS with fixed 
periodicity. The simulation results are provided in Table \ref{tab:stopping distance} considering different possible
road surfaces. 
\begin{table}[h] \vspace{-.4cm} \scriptsize
    \caption{Stopping Distance in Kilometer}
    \label{tab:stopping distance}
\resizebox{\linewidth}{!}{
    \begin{tabular}{|p{1.7cm}|p{1.2cm}|p{1.2cm}|p{1.2cm}|p{0.5cm}|p{0.5cm}|} \hline
      \centering ABS&\multicolumn{5}{c|}{Road Surface} \\
      \cline{2-6}
       \centering Controller&\centering Dry\par Asphalt& \centering Gravel& \centering Loose\par Gravel&\multicolumn{2}{c|}{Wet}\\ \hline
      \centering Existing model&\centering 3.073&\centering 3.424&\centering
3.771&\multicolumn{2}{c|}{4.269}\\ \hline
      \centering Multi-mode&\centering 3.080&\centering 3.433&\centering 3.795&\multicolumn{2}{c|}{4.289}\\ \hline
    \end{tabular}} \vspace{-.3cm}
  \end{table}
We observe that the stopping distance is nearly same for both the controllers. For this `panic braking' scenario, the
estimated percentage of time spent in each of $E$, $N1$ and $N0$ modes is highlighted in Fig.
\ref{fig:ECU_comparision}. 
It is evident from Fig. \ref{fig:ECU_comparision} that we can save significant amount of ECU
bandwidth (30\% - 50\%) using a multi-mode controller as compared to the controller with fixed periodicity. 

\par
\begin{figure}[h]
  \centering \vspace{-.25cm}
   \epsfig{file=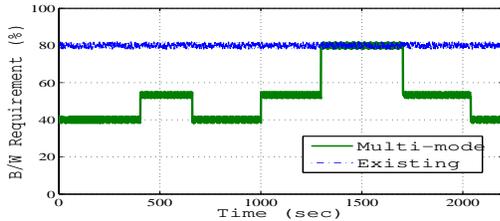, height=1.2in, width=3in} \vspace{-.4cm}
   \caption{ECU, General Cruising Scenario} \vspace{-.25cm}
    \label{fig:ECU}
\end{figure}

 Further, we investigate the utility of our multi-mode ABS controller in a general cruising scenario where the car
is being driven in a speed range of $0km/h$ to $200km/h$, in various traffic densities. The ECU bandwidth 
requirement for this scenario is shown in Fig. \ref{fig:ECU}. It is evident that the multi-mode controller
requires much lesser bandwidth compared to the existing controller since the supervisory automaton schedules the
controller in the infrequent sampling modes for most of the time as described in Section \ref{automata}. In that way, we
can guarantee a significant amount of bandwidth saving which may be utilized for scheduling other tasks. 
%
\par
\begin{figure}[h]
  \centering \vspace{-.2cm}
   \epsfig{file=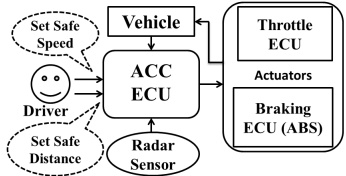,scale=0.2} \vspace{-.4cm}
    \caption{ACC System Overview~\cite{bosch_safety}} \vspace{-.25cm}
    \label{fig:acc}
\end{figure}

 We further demonstrate the promise of multi-mode sampling for a multiple ECU scenario taking an Adaptive Cruise
Control (ACC) system as a running example. ACC is an automobile safety critical driver assistance system which
automatically adjusts the vehicle speed in order to maintain a safe distance from vehicles ahead. In case of an ACC 
system, the driver sets a safe cruising speed and a desired safe distance (from preceding vehicle) as controller
inputs. The other inputs of an ACC controller which comes from the radar sensor are preceding vehicle speed and 
approximate distance from preceding vehicle as shown in Fig. \ref{fig:acc}. 
\par ACC is a drive assist system for highway cruising which monitors the inputs and decides cruising speed or distance
to lead vehicle. When the applied brake pedal pressure is high, the ACC is overridden by the braking
controller (ABS). The operating modes of ACC controller can be classified into `active', `suspended' and `idle' while our 
multi-mode ABS operates in three different modes as discussed previously. Generally, the ACC system and the braking 
controller are mapped to separate ECUs. We can achieve significant reduction in bandwidth requirement by sharing an ECU
between these features. 

\par
\begin{figure}[h]
  \centering \vspace{-.2cm}
   \epsfig{file=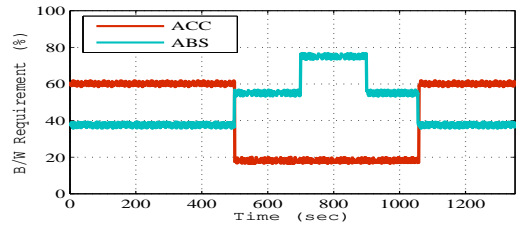, height=1.2in, width=3in} \vspace{-.35cm}
   \caption{ECU Sharing: ACC-ABS} \vspace{-.22cm}
    \label{fig:ACC_ABS_BW}
\end{figure}

 When the ACC is suspended, because the applied brake pedal pressure is high, we schedule the ABS controller with 
frequent sampling rate and the ACC controller with relatively infrequent sampling rates. On the other hand, when the 
ACC is active, i.e. applied brake pedal pressure is low or null, then the ACC controller is scheduled with 
frequent sampling rate and the ABS controller is scheduled with relatively infrequent sampling rate. In Fig.
\ref{fig:ACC_ABS_BW}, we show the bandwidth requirement of both ACC and ABS controllers when scheduled in a single ECU
while providing satisfactory control performance.
\section{Conclusions}
The present work provides a methodology for adaptively regulating the sampling rate of embedded software based controllers leading to significant reduction of computational
resource requirement. Applying the methodology on single controller based systems like ABS has shown that 30\% - 50\% reduction in ECU bandwidth requirement is possible. 
Further, it was also shown that the method smoothly scales up for multiple controller based systems. Our future research shall focus on giving a sound formal underpinning to 
the method of creating multiple sampling modes for software based  controllers and creating a tool flow which mechanizes the synthesis of such multi-mode controllers.

\bibliographystyle{abbrv}
\bibliography{sigproc}

\begin{thebibliography}{10}

\bibitem{bosch_safety}
{\em Safety, Comfort and Convenience Systems}.
\newblock Robert Bosch GmbH, 2006.

\bibitem{discretecontrolbook}
K.~{\AA}str{\"o}m and B.~Wittenmark.
\newblock {\em Computer controlled systems: theory and design}.
\newblock Prentice Hall, 1984.

\bibitem{Buttazzo}
G.~Buttazzo.
\newblock Research trends in real-time computing for embedded systems.
\newblock {\em SIGBED Rev.}, 3(3).

\bibitem{cervin2011paper}
A.~Cervin, M.~Velasco, P.~Marti, and A.~Camacho.
\newblock Optimal online sampling period assignment: Theory and experiments.
\newblock {\em IEEE Transactions on Control Systems Technology}, 19(4), 2011.

\bibitem{henrikson}
D.~Henriksson and A.~Cervin.
\newblock Optimal on-line sampling period assignment for real-time control
  tasks based on plant state information.
\newblock In {\em 44th IEEE CDC-ECC}, 2005.

\bibitem{khalil1992nonlinear}
H.~Khalil.
\newblock {\em Nonlinear systems}.
\newblock Macmillan Pub.Co., 1992.

\bibitem{kuo_book}
B.~Kuo.
\newblock {\em Automatic control systems}.
\newblock Prentice Hall, 1962.

\bibitem{ABS_paper}
M.~Schinkel and K.~Hunt.
\newblock Anti-lock braking control using a sliding mode like approach.
\newblock In {\em Proceedings of the ACC,}, volume~3, 2002.

\bibitem{seto}
D.~Seto, J.~Lehoczky, L.~Sha, and K.~Shin.
\newblock On task schedulability in real-time control systems.
\newblock In {\em 17th IEEE RTSS}, 1996.

\bibitem{wong2001theory}
J.~Wong.
\newblock {\em Theory of Ground Vehicles}.
\newblock Wiley, 2001.

\end{thebibliography}
\newpage
\appendix
\section{Basic Control System}\label{A:Control}
\begin{figure}[h]
 \centering
 \epsfig{file=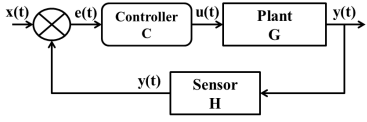, scale=0.22}
    \caption{Continuous Time Control System}
    \label{fig:Continious Controller}
\end{figure}

\begin{figure}[h]
 \centering
 \epsfig{file=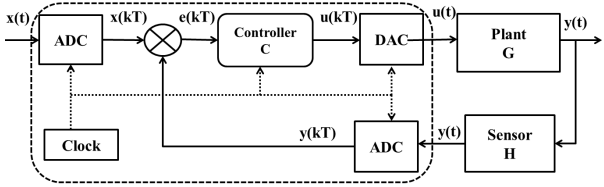, scale=0.22}
 \caption{Discrete Time Control System}
    \label{fig:DigitalController}
\end{figure}

Any continuous feedback control system\cite{kuo_book} can be represented as shown in Fig. \ref{fig:Continious Controller}, where $x(t)$ is the input
to the system, $e(t)$ is the error signal, $u(t)$ is the controller output and $y(t)$ is the plant output fed back to the controller using a sensor(H).
Correspondingly, a Discrete time Feedback Control System can be represented as shown in Fig. \ref{fig:DigitalController},
where the dotted box highlights the discretized controller \cite{discretecontrolbook}.
Generally, the sampling of the continuous signal is done at a constant rate 
$T$, which is known as the sampling period or interval. The sampled signal $y_k=y(kT)$ is the discretized signal with 
$k\in\mathbb{N}$. For understanding the design of a discretized control system, let us first consider a simple
continuous domain transfer function for the controller C in Fig. \ref{fig:Continious Controller} given as follows.
\begin{center}$C(s) = \frac{U(s)}{E(s)} = \frac{K(s+a)}{(s+b)}$\end{center}
The corresponding time domain representation shall be,
\begin{center}$\frac{du}{dt} + bu = K(\frac{de}{dt}+ae)$\end{center}
The Euler's approximation of the first order derivative is represented as:
\begin{equation}
 \label{eq:euler}
\frac{dx}{dt} \approx \frac{x_{k+1} - x_k}{\Delta t}\\
\end{equation}
Applying Euler's approximation of first order derivative \cite{discretecontrolbook} on the continuous differential
equation, we get the following discrete difference equation.
\begin{center}$\frac{u_{k+1}-u_k}{\Delta t} + bu_k = K(\frac{e_{k+1}-e_k}{\Delta t} + ae_k)$\end{center}
Generally, $\Delta t$, K, a and b are fixed. The digital controller updates the control signal every cycle as per the
following equation,
\begin{equation}
 \label{eq:1}
u_{k+1} = -a_1u_k + b_0e_{k+1} + b_1e_k\\
\end{equation}
where, $b_0 = K$, $b_1 = K(a \Delta t - 1)$, $a_1 = (b \Delta t - 1)$. In general, for any transfer function,
the control signal output depends on $n$ previous control signal output instances and $m$ previous error signal
instances which can be represented as\cite{discretecontrolbook},
\begin{equation}
 \label{eq:generalized 1}
  \begin{split}
    u_k &= -a_1u_{k-1} -a_2u_{k-2} - \dots - a_nu_{k-n} + b_0e_{k} +\\
	& b_1e_{k-1} + b_2e_{k-2} + \dots + b_me_{k-m}\\
  \end{split}
\end{equation}
The discrete signals $u_{k-1}$,$u_{k-2}$, \dots are the delayed versions of $u_k$ by sampling period $T, 2T,$\dots respectively.
\subsection{Stability: Discrete Time Control System}
\begin{figure}[h]
 \centering \vspace{-.3cm}
 \epsfig{file=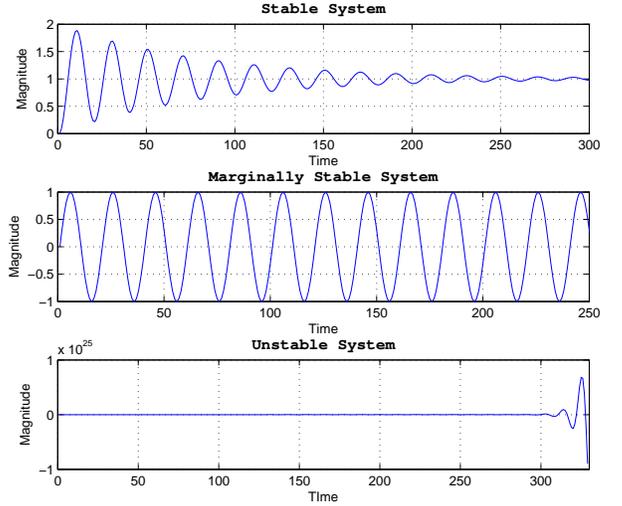,scale=0.5} \vspace{-.9cm}
 \caption{Impulse Response and Stability} \vspace{-.1cm}
\label{fig:impusle_stability}
\end{figure}

\begin{figure}[h]
 \centering
 \epsfig{file=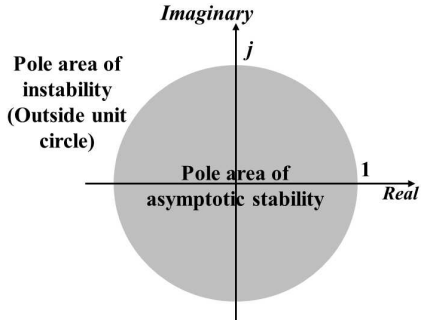, scale=0.3}
 \caption{Unit Circle: Stability areas in Complex Plane}
\label{fig:unit-circle}
\end{figure}

The stability property of this system can be defined from the impulse response of a system as
\begin{itemize}
 \item Asymptotic stable system: The steady state impulse response is zero.\vspace{-.8cm}
\begin{center}
 $$\lim_{k \to \infty} yδ(k) = 0 $$
\end{center}
 \item Marginally stable system: The steady state impulse response is different from zero, but limited.\vspace{-.8cm}
\begin{center}
 $$\lim_{k \to \infty} 0< yδ(k) < \infty$$
\end{center}
 \item Unstable system: The steady state impulse response is unlimited.\vspace{-.9cm}
\begin{center}
 $$\lim_{k \to \infty} yδ(k) = \infty$$
\end{center}
\end{itemize}
where $y(k)$ is the impulse response of the corresponding system. The impulse response for different stability property
is illustrated in Fig. \ref{fig:impusle_stability}.
Let us assume a control system with input $u$ and output $y$. The transfer function of any discrete time control system
can be represented as
\begin{center}$C(z) = \frac{y(z)}{u(z)} = \frac{bz}{(z-p)}$\end{center}
where $p$ is the pole which is in general a complex number and can be written in polar from as $p=me^{j\theta}$ where 
$m$ is the magnitude and $\theta$ is the phase. The impulse response of the system can be given as
\begin{center}$y(k) = Z^{-1}\{\frac{bz}{z-p}\} = b\lvert m\lvert^ke^{jk\theta}$\end{center}
Thus, it is the magnitude $m$ which determines if the steady state impulse response converges towards zero or not. The
relationship between stability and pole placement can be stated as follows.
\begin{itemize}
 \item Asymptotic stable system: All poles lie inside (none is on) the unit circle, or what is the same: all poles have magnitude less than 1.
 \item Marginally stable system: One or more poles but no multiple poles are on the unit circle.
 \item Unstable system: At least one pole is outside the unit circle.
\end{itemize}
%
The situation is graphically shown in Fig. \ref{fig:unit-circle}.
\section{Quarter Vehicle Modeling}\label{A:0}
\begin{figure}[h]
 \centering
 \epsfig{file=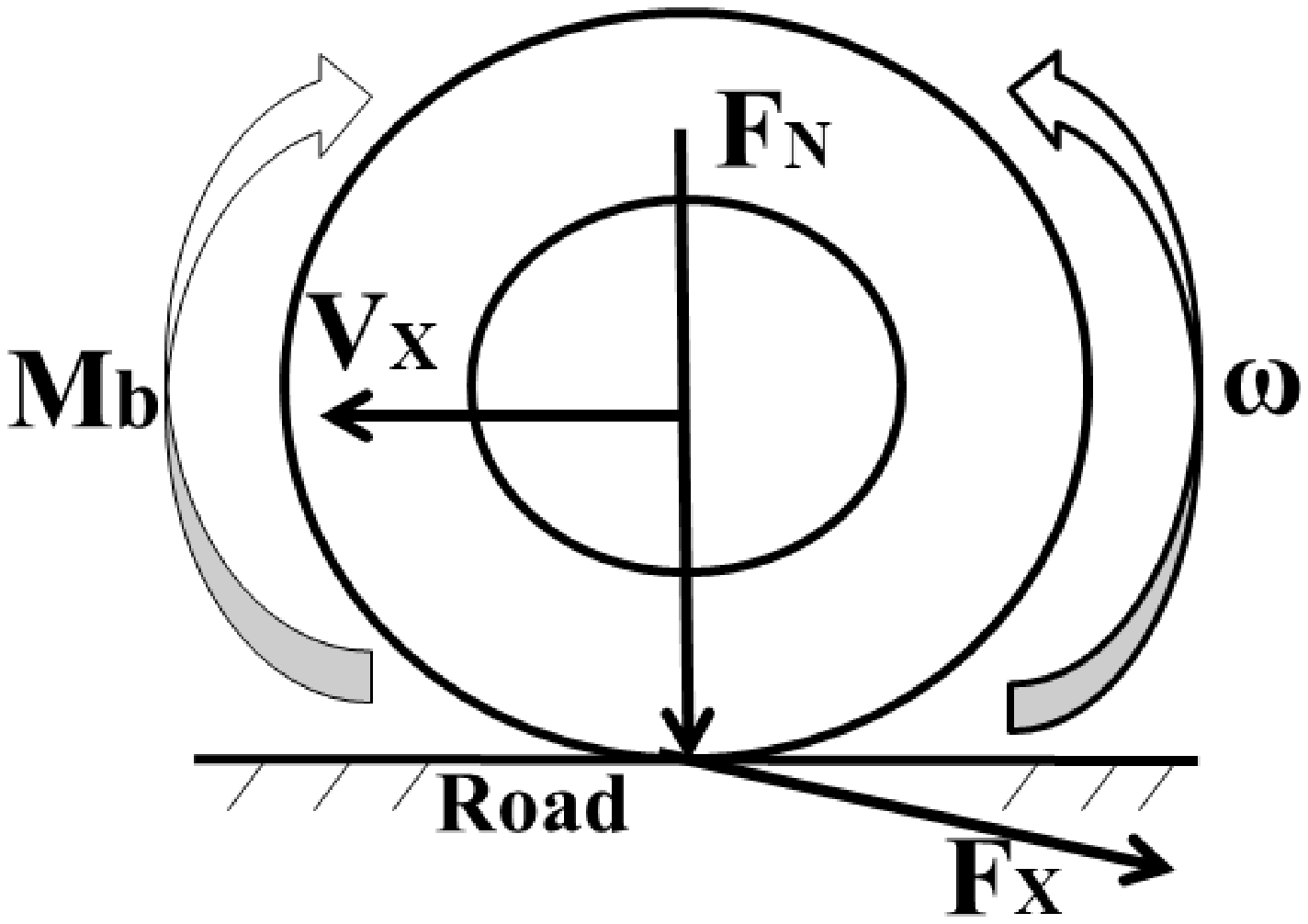, scale=0.3}
 \caption{1/4 car forces and torques}
\label{fig:1/4car}
\end{figure}

A quarter car model is shown in figure \ref{fig:1/4car}.
Here, m is the mass of the quarter vehicle, $V_x$ is lateral speed of the vehicle,
$\omega$ is the angular speed of the wheel, $F_N$ is the vehicle vertical force, $F_x$ is tire frictional force,
$M_b$ is the braking torque, R is the wheel radius and $J_\omega$ is the wheel inertia.
Wheel slip $\lambda$ is represented as:
\begin{center}
  $\lambda = 1 - \frac{\omega R}{V_x}$
\end{center}
The relationship between $F_N$ and $F_x$\cite{bosch_safety} is given as:
\begin{center}
 $F_x = -\mu(\lambda)F_N$
\end{center}
where $\mu(\lambda)$ is the frictional coefficient of the road surface.
The non-linear equations for designing a quarter car model can be given as\cite{ABS_paper}:
\begin{center}
  $\dot{V_x} = -\frac{1}{m}F_N\mu(\lambda)$ \\

  $\dot{\omega} = \frac{R}{J_\omega}F_N\mu(\lambda) - \frac{M_b}{J_\omega}$ \\

  $\dot{\lambda} = \frac{\dot{V_x}(1-\lambda)-\dot\omega R}{V_x}$ \\

  $\dot{\lambda} = -\frac{1}{V_x}[\frac{1}{m}(1-\lambda)+\frac{R^2}{J_\omega}]F_N\mu(\lambda) + \frac{1}{V_x}\frac{R}{J_\omega}M_b$\\
\end{center}

For linearizing the system approximation using the Taylor series expansion can be expressed as \cite{khalil1992nonlinear}: 
$f(\lambda,V_x)\approx$ $f(\lambda',V'_x)$ + $\frac{df}{d\lambda}|_{\lambda',V'_x}$ $(\lambda - \lambda')$ + 
$\frac{df}{dV}|_{\lambda',V'_x}$ $(V -V'_x)$. From this from this we can derive a linear (affine) system description
as:
\begin{center}
    $\dot{x} = A_l x + E_l + B_l u^*$ \\
     $y = C_l x + D_l u^*$ \\
     $l =f(x)$ \\
\end{center}

where, $A_l$, $B_l$, $C_l$, $D_l$ are system input and output matrices respectively.
$E_l$ are the affine terms and $f(x)$ is the function telling the validity of linearizion and $x^T = [V_x,\lambda]$
, $u^* = M_b \cdot V_x$ , $y = [\lambda]$. From Fig. \ref{fig:mu-slip} the relationship between wheel slip and
frictional coefficient can be approximated by using piecewise linear function as,
\begin{equation}
 \label{eq:A_mu}
  \mu (\lambda) =
  \begin{cases}
     \alpha \lambda ,& \lambda \leq 0.2 \\
     - \frac{1}{2} \lambda +\frac{3}{4} + \beta, & \lambda > 0.2
  \end{cases}
\end{equation}
where, $\alpha$ $\in$ $[4.8, 5.1, 5.46, 6.4]$ and $\beta\in [-0.1,0.1]$.
\begin{equation}\label{a<0.2}
 A_l = \begin{bmatrix} 0 & -\alpha\frac{F_N}{m} \\ \alpha\frac{F_NR^2\lambda'}{V'^2_xJ_\omega} & \alpha\frac{F_NR^2}{V'_xJ_\omega} \end{bmatrix}
\end{equation}
\begin{equation}\label{e<0.2}
 E_l = \begin{bmatrix} 0 \\ -\alpha\frac{F_NR^2\lambda'}{V'_xJ_\omega} \end{bmatrix}
\end{equation}

and for $\lambda>0.2$
\begin{equation}\label{a>0.2}
 A_l = \begin{bmatrix} 0 & \frac{F_N}{4m} \\ (-\frac{\lambda'}{2}+\frac{3}{4})\frac{F_NR^2}{V'^2_xJ_\omega}\pm 0.1\frac{F_NR^2}{V'^2_xJ_\omega} & \frac{F_NR^2}{4V'_xJ_\omega} \end{bmatrix}
\end{equation}
\begin{equation}\label{e>0.2}
 E_l = \begin{bmatrix} (-\frac{3}{4}\pm0.1)\frac{F_N}{m} \\ (\frac{\lambda'}{2}-\frac{3}{2})\frac{F_NR^2}{V'_xJ_\omega}\pm 0.2\frac{F_NR^2}{V'_xJ_\omega} \end{bmatrix}
\end{equation}
The stability analysis of this system can be done using Lyapunov, Bode, Nyquist, Unit Circle or Hurwitz stability criteria \cite{ABS_paper}.
\section{Stability v/s Sampling Period}
In this section we provide few examples relating to the variation in stability with respect to sampling periods.
\subsection{Simple Examples:} \label{A:1}
\begin{figure}[h]
 \centering
 \epsfig{file=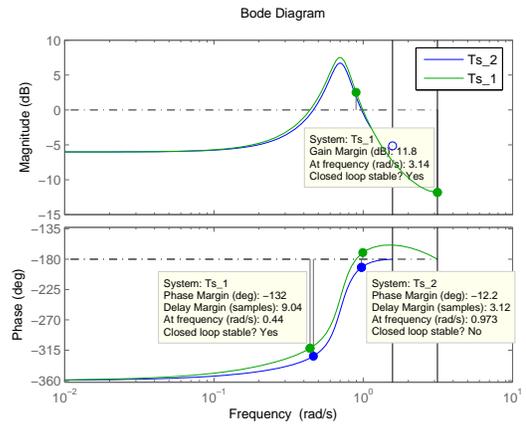, scale=0.5}
 \caption{$T_s = 2$s, Unstable; $T_s = 1$s, Stable}
    \label{fig:simple_bode1}
\end{figure}

Let the transfer function of any arbitrary system be,
\begin{center}
     $U(s)=  \frac{s+0.5}{ms^2+bs+u}$
\end{center}
\textbf{Case 1:} Let m=2, b=-0.5, u=1. For $T_s$=2s the system unstable but for $T_s$=1s the system is stable.
The values of m, b, u are same for both of the sampling intervals. The corresponding stability response is shown in Fig. \ref{fig:simple_bode1}.

\begin{figure}[h]
 \centering \vspace{-.3cm}
 \epsfig{file=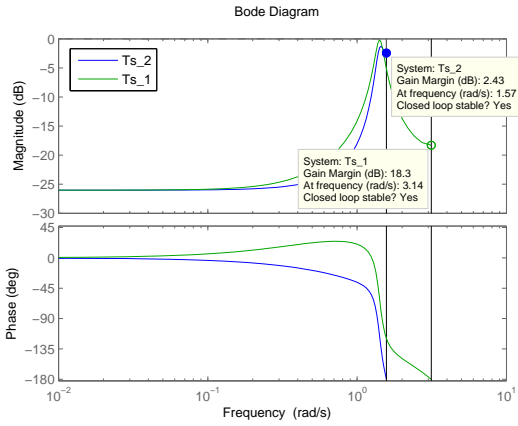, scale=0.5} \vspace{-.4cm}
 \caption{$T_s = 2$s, and $T_s = 1$s, Both Stable}\vspace{-.3cm}
    \label{fig:simple_bode2}
\end{figure}

\textbf{Case 2:} Let m=5, b=1, u=10. In this case the system stable for both the sampling intervals $T_s$=2s and $T_s$=1s.
The values of m, b, u are same for both of the sampling intervals. The corresponding stability response is shown in Fig. \ref{fig:simple_bode2}.
\subsection{ABS example}\label{A:2}
We provide more examples of stabiliy variation with sampling interval for the ABS controller model.

\begin{figure}[h]
 \centering \vspace{-.3cm}
 \epsfig{file=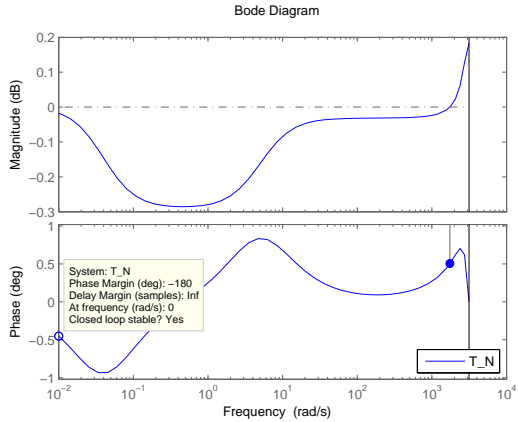, scale=0.5}\vspace{-.4cm}
 \caption{$T_s = 0.6$ ms, Stable} \vspace{-.3cm}
    \label{fig:abs_bode0}
\end{figure}

\textbf{Case 1:} For V = 10 Km/h, $\lambda$ = 0.1, the stability response is shown in Fig.
\ref{fig:abs_bode0}.

\begin{figure}[h]
 \centering \vspace{-.3cm}
 \epsfig{file=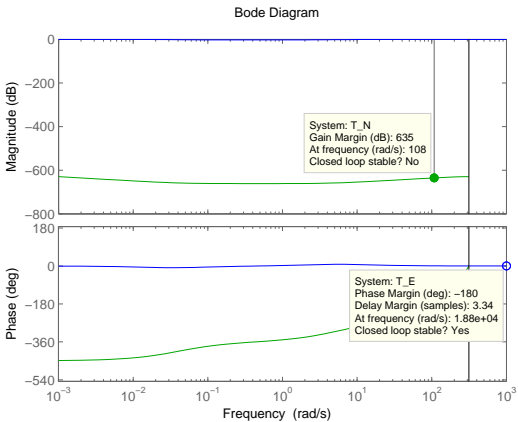, scale=0.5} \vspace{-.4cm}
 \caption{$T_s = 0.6$ ms, Unstable; $T_s = 0.1$ ms, Stable}
    \label{fig:abs_bode1} \vspace{-.6cm}
\end{figure}

\begin{figure}[h]
 \centering \vspace{-.2cm}
 \epsfig{file=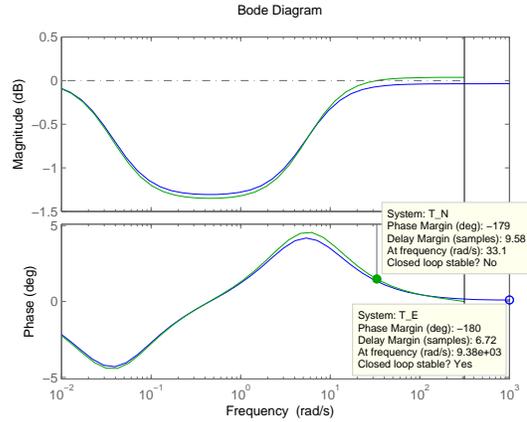, scale=0.5} \vspace{-.4cm}
 \caption{$T_s = 0.6$ ms, Unstable; $T_s = 0.3$ ms, Stable}
    \label{fig:abs_bode2}
\end{figure}

\textbf{Case 2:} For V = 100 Km/h, $\lambda$ = 0.6, the stability response is shown in Fig.
\ref{fig:abs_bode1}.
\newpage
\textbf{Case 3:} For V = 40 Km/h, $\lambda$ = 0.2, the stability response is shown in Fig.
\ref{fig:abs_bode2}.

\begin{figure}[h]
 \centering \vspace{-.2cm}
 \epsfig{file=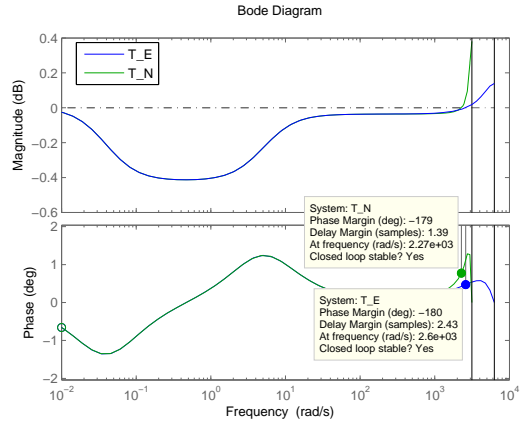, scale=0.5} \vspace{-.4cm}
 \caption{$T_s = 0.6$ ms, and $T_s = 0.1$ ms, Both Stable}
    \label{fig:abs_bode3}
\end{figure}

\textbf{Case 4:} For V = 15 Km/h, $\lambda$ = 0.3, the stability response is shown in Fig.
\ref{fig:abs_bode3}.
\par From these examples we can observe that sampling period has a major role to play in deciding the stability of a
software based controller.
\end{document}